\documentclass{article}

\usepackage{fullpage,latexsym}
\usepackage{epsfig}
\usepackage{pslatex}

\begin{document}

\title{Algorithmic Clustering of Music}
\author{Rudi Cilibrasi\thanks{Supported in part by NWO.
Address:
CWI, Kruislaan 413,
1098 SJ Amsterdam, The Netherlands.
Email: {\tt Rudi.Cilibrasi@cwi.nl}.}
\\CWI
\and
Paul Vitanyi\thanks{Supported in part by the
EU project RESQ, IST--2001--37559, the NoE QUIPROCONE
+IST--1999--29064,
the ESF QiT Programmme, and the EU Fourth Framework BRA
 NeuroCOLT II Working Group
EP 27150. 
Address:
CWI, Kruislaan 413,
1098 SJ Amsterdam, The Netherlands.
Email: {\tt Paul.Vitanyi@cwi.nl}.}\\CWI and University of Amsterdam
\and
Ronald de Wolf\thanks{Supported in part by EU project RESQ, IST-2001-37559.
Address:
CWI, Kruislaan 413,
1098 SJ Amsterdam, The Netherlands.
Email: {\tt Ronald.de.Wolf@cwi.nl}.}\\
CWI}
\date{}
\maketitle

\begin{abstract}
We present a fully automatic method for music classification,
based only on compression of strings that represent the music
pieces. The method uses no background knowledge 
about music whatsoever: it is completely
general and can, without change, be used in different areas
like linguistic classification and genomics. It is based on an ideal
theory of the information content in individual objects
(Kolmogorov complexity), information distance, and a 
universal similarity metric. Experiments show that the method
distinguishes reasonably well between various musical genres
and can even cluster pieces by composer.
\end{abstract}

\section{Introduction}

All musical pieces are similar, but some are more similar than others.
Apart from being an infinite source of discussion (``Haydn is just 
like Mozart --- no, he's not!''), such similarities are also 
crucial for the design of efficient music information retrieval systems. 
The amount of digitized music available on the internet has grown 
dramatically in recent years, both in the public domain 
and on commercial sites. Napster and its clones are prime examples.
Websites offering musical content in some form or other
(MP3, MIDI, \ldots) need a way to organize their wealth of material;
they need to somehow classify their files according to 
musical genres and subgenres, putting similar pieces together.
The purpose of such organization is to enable users 
to navigate to pieces of music they already know and like, 
but also to give them advice and recommendations
(``If you like this, you might also like\ldots'').
Currently, such organization is mostly done manually by humans,
but some recent research has been looking into the possibilities 
of automating music classification.

A human expert, comparing different pieces of music with the aim to cluster
likes together, will generally look for certain specific similarities.
Previous attempts to automate this process do the same.
Generally speaking, they take a file containing a piece of music
and extract from it various specific numerical features,
related to pitch, rhythm, harmony etc.  
One can extract such features using for instance 
Fourier transforms~\cite{TC02} or wavelet transforms~\cite{GKCwavelet}.
The feature vectors corresponding to the various files are then 
classified or clustered using existing classification software, based on 
various standard statistical pattern recognition classifiers~\cite{TC02}, 
Bayesian classifiers~\cite{DTWml},
hidden Markov models~\cite{CVfolk},
ensembles of nearest-neighbor classifiers~\cite{GKCwavelet}
or neural networks~\cite{DTWml,Sneural}.
For example, one feature would be to look for rhythm in the sense 
of beats per minute. One can make a histogram where each histogram 
bin corresponds to a particular tempo in beats-per-minute and 
the associated peak shows how frequent and strong that
particular periodicity was over the entire piece. In \cite{TC02}
we see a gradual change from a few high peaks to many low and spread-out
ones going from hip-hip, rock, jazz, to classical. One can use this
similarity type to try to cluster pieces in these categories.
However, such a method requires specific and detailed knowledge of 
the problem area, since one needs to know what features to look for. 

Our aim is much more general.
We do not look for similarity in specific features known to 
be relevant for classifying music; 
instead we apply a general mathematical theory of similarity.
The aim is to capture, in a single similarity metric, 
{\em every effective metric\/}: 
effective versions of Hamming distance, Euclidean distance, 
edit distances, Lempel-Ziv distance, and so on.
Such a metric would be able to simultaneously detect {\em all\/}
similarities between pieces that other effective metrics can detect.
Rather surprisingly, such a ``universal'' metric indeed exists.
It was developed in \cite{LBCKKZ01,Li01,Li03}, based on the 
``information distance'' of \cite{LiVi97,BGLVZ98}. 
Roughly speaking, two objects are deemed close if
we can significantly ``compress'' one given the information
in the other, the idea being that if two pieces are more similar,
then we can more succinctly describe one given the other.
Here compression is based on the ideal mathematical notion of Kolmogorov
complexity, which unfortunately is not effectively computable.
It is well known that when a pure mathematical theory 
is applied to the real world, for example in hydrodynamics
or in physics in general, we can in applications only approximate
the theoretical ideal. But still the theory gives a framework and foundation
for the applied science. Similarly here. We replace the ideal but 
noncomputable Kolmogorov-based version by standard compression techniques.
We lose theoretical optimality in some cases, but gain an efficiently 
computable similarity metric intended to
 approximate the theoretical ideal.
In contrast, a later and partially independent
compression-based approach of
\cite{BCL02a,BCL02b} for building language-trees---while 
citing \cite{LiVi97,BGLVZ98}---is by {\em ad hoc\/} arguments 
about empirical Shannon entropy and Kullback-Leibler distance 
resulting in non-metric distances.  

Earlier research has demonstrated that this new universal similarity 
metric works well on concrete examples in very different application
fields---the first completely automatic construction
of the phylogeny tree based on whole mitochondrial genomes,
\cite{LBCKKZ01,Li01,Li03} and
a completely automatic construction of a language tree for over 50
Euro-Asian languages \cite{Li03}. 
Other applications, not reported in print, 
are detecting plagiarism in student programming assignments 
\cite{SID}, and phylogeny of chain letters.

In this paper we apply this compression-based method to the classification of 
pieces of music. We perform various experiments on sets of 
mostly classical pieces given as MIDI (Musical Instrument Digital 
Interface) files. This contrasts with most earlier research,
where the music was digitized in some wave format or other
(the only other research based on MIDI that we are aware 
of is~\cite{DTWml}).
We compute the distances between all pairs of pieces, 
and then build a tree containing those pieces in a way that 
is consistent with those distances. 
First, as proof of principle, we run the program on three
artificially generated data sets, where we know what 
the final answer should be.
The program indeed classifies these perfectly.
Secondly, we show that our program can distinguish between various
musical genres (classical, jazz, rock) quite well.
Thirdly, we experiment with various sets of classical pieces.
The results are quite good (in the sense of conforming 
to our expectations) for small sets of data,
but tend to get a bit worse for large sets.
Considering the fact that the method knows nothing
about music, or, indeed, about any of the other areas
we have applied it to elsewhere, one is reminded of Dr Johnson's
remark  
about a dog's walking on his hind legs: 
``It is not done well; but you are surprised to find it done at all.''

The paper is organized as follows.
We first give a  domain-independent overview of compression-based
clustering: the ideal distance metric based on Kolmogorov complexity,
and the quartet method that turns the matrix of distances into a tree.
In Section~\ref{secdetails} we give the details of the current application
to music, the specific file formats used etc.
In Section~\ref{secresults} we report the results of our experiments.
We end with some directions for future research.

\section{Algorithmic Clustering}

\subsection{Kolmogorov complexity}
Each object (in the application of this paper: each piece of music) is
coded as a string $x$ over a finite alphabet, say the binary
alphabet. 
The integer $K(x)$ gives
the length of the shortest compressed binary version from which
$x$ can be fully reproduced, 
also known as the {\em Kolmogorov complexity\/} of $x$. 
``Shortest'' means the minimum taken over every
possible decompression program, the
ones that are currently known as well as the ones that are possible
but currently unknown. We explicitly write only ``decompression''
because we do not even require that there is also a program that
compresses the original file to this compressed version---if there
is such a program then so much the better.  
Technically, the definition of Kolmogorov complexity is as follows.
First, we fix a syntax for expressing all and only computations (computable
functions). This can be in the form of an enumeration of all 
Turing machines, but also an enumeration of all syntactically correct
programs in some universal programming language like Java, Lisp, or C.
We then define the Kolmogorov complexity of a finite binary string
as the length of the shortest Turing machine, Java program, etc.
in our chosen syntax. Which syntax we take is unimportant, but
we have to stick to our choice. This choice attaches a definite positive
integer as the Kolmogorov complexity to each finite string.

Though defined in terms of a
particular machine model, the Kolmogorov complexity
is machine-independent up to an additive
constant
 and acquires an asymptotically universal and absolute character
through Church's thesis, and from the ability of universal machines to
simulate one another and execute any effective process.
  The Kolmogorov complexity of an object can be viewed as an absolute
and objective quantification of the amount of information in it.
   This leads to a theory of {\em absolute} information {\em contents}
of {\em individual} objects in contrast to classic information theory
which deals with {\em average} information {\em to communicate}
objects produced by a {\em random source}.

So $K(x)$ gives the length of the ultimate 
compressed version, say $x^*$, of $x$. 
This can be considered as the amount of information, number of bits,
contained in the string. Similarly, $K(x|y)$ is the minimal number of
bits (which we may think of as constituting a computer program) 
required to reconstruct $x$ from $y$.
In a way $K(x)$ expresses the individual ``entropy'' of $x$---the
minimal number of bits to communicate $x$ when sender and
receiver have no knowledge where $x$ comes from. For example,
to communicate Mozart's ``Zauberfl\"ote'' from a library of a 
million items requires at most 20 bits ($2^{20}\approx 1,000,000$), 
but to communicate it from scratch requires megabits.
For more details on this pristine notion of individual
information content we refer to the textbook
\cite{LiVi97}.

\subsection{Distance-based classification}

As mentioned, our approach is based on a new 
very general similarity distance, classifying the objects in
clusters of objects that are close together according to this distance.
In mathematics, lots of different distances arise in all sorts of contexts,
and one usually requires these to be a `metric', since otherwise 
undesirable effects may occur. 
A metric is a distance function $D(\cdot,\cdot)$ that assigns
a non-negative distance $D(a,b)$ to any two objects $a$ and $b$, in such a way that 
\begin{enumerate}
\item $D(a,b)=0$ only where $a=b$ 
\item $D(a,b)=D(b,a)$ (symmetry)
\item $D(a,b)\leq D(a,c)+D(c,b)$ (triangle inequality)
\end{enumerate}
A familiar example of a metric is the Euclidean metric, 
the everyday distance $e(a,b)$ between two objects $a,b$ 
expressed in, say, meters.  
Clearly, this distance satisfies the properties
$e(a,a)=0$, $e(a,b)=e(b,a)$, and $e(a,b) \leq e(a,c) + e(c,b)$
(Substitute $a=$ Amsterdam, $b=$ Brussels, and $c=$ Chicago.) 
We are interested in ``similarity metrics''. 
For example, if the objects are classical music pieces
then the function $D(a,b)=0$ if $a$ and $b$ are by the same composer
and $D(a,b)=1$ otherwise, is a similarity metric, albeit a somewhat elusive one. 
This captures only one, but quite a significant, similarity aspect 
between music pieces. 

In \cite{Li03}, a new theoretical approach
to a wide class of similarity metrics was proposed:
the ``normalized information distance'' is a metric, and it is 
universal in the sense that this single metric uncovers all similarities 
simultaneously that the metrics in the class uncover separately.
This should be understood in the sense that if two pieces of music
are similar (that is, close) according to the particular feature described by 
a particular metric, then they are also similar (that is, close)
in the sense of the normalized information distance metric. This justifies
calling the latter {\em the\/} similarity metric.
Oblivious to the problem area concerned, simply using the distances
according to the similarity metric, our method fully automatically
classifies the objects concerned, be they music pieces, 
text corpora, or genomic data. 

More precisely, the approach is as follows.
Each pair of such strings $x$ and $y$ is assigned a distance
\begin{equation}\label{eq.distance}
d(x,y) = \frac{\max\{K(x|y),K(y|x)\}}{\max\{K(x),K(y) \}} .
\end{equation}
There is a natural interpretation to $d(x,y)$: If, say, $K(y) \geq K(x)$
then we can rewrite
\[d(x,y) = \frac{K(y)-I(x:y)}{K(y)} , \]
where $I(x:y)$ is the information in $y$ about $x$ satisfying
the symmetry property $I(x:y)=I(y:x)$ up to a logarithmic additive error
\cite{LiVi97}.
That is, the distance $d(x,y)$ between $x$ and $y$ is the
number of bits of information that is not shared between the two strings
per bit of information that could be maximally shared between the two strings.

It is clear that $d(x,y)$ is symmetric, and in \cite{Li03} it
is shown that it is indeed a metric. Moreover, it is universal
in the sense that every metric expressing some similarity
that can be computed from the objects concerned is comprised
(in the sense of minorized) by $d(x,y)$. It is these distances that we
will use, albeit in the form of a rough approximation: for
$K(x)$ we simply use standard compression software like `gzip', `bzip2', or
`compress'. To compute the conditional version, $K(x|y)$ we use
a sophisticated theorem, known as ``symmetry of algorithmic information''
in \cite{LiVi97}. This says 
\begin{equation}\label{eq.condition}
K(y|x) \approx K(xy)-K(x), 
\end{equation}
so to compute the conditional complexity $K(x|y)$ we can just take
the difference of the unconditional complexities $K(xy)$ and $K(y)$.
This allows us to approximate $d(x,y)$ for every pair $x,y$.

Our actual practice falls short of the ideal theory in at least 
three respects:

(i) The claimed universality of the similarity distance $d(x,y)$
holds only for indefinitely long sequences $x,y$. Once we consider
strings $x,y$ of definite length $n$, the similarity distance
is only universal with respect to ``simple'' computable normalized information
distances, where ``simple'' means that they are computable by programs
of length, say, logarithmic or polylogarithmic in $n$.
This reflects the fact that, technically speaking, the universality 
is achieved by summing the weighted contribution of all
similarity distances in the class considered with respect
to the objects considered. Only similarity distances of which
the complexity is small (which means that the weight is large)
with respect to the size of the data concerned kick in. 

(ii) The Kolmogorov complexity is not computable, and it is 
in principle impossible to compute how far off our approximation
is from the target value in any useful sense.

(iii) To approximate the information distance in a practical sense
we use the standard compression program bzip2. While better compression
of a string will always  approximate the Kolmogorov complexity better,
this is, regrettably, not true for the (normalized) information distance.
Namely, using (\ref{eq.condition}) we consider the difference of
two compressed quantities. Different compressors may compress
the two quantities differently, causing an increase in the
difference even when both quantities are compressed better (but
not both as well). In the normalized information distance we
also have to deal with a ratio that causes the same problem.
Thus, a better compression program may not necessarily mean
that we also approximate the (normalized) information distance
better. This was borne out by the results of our experiments using
different compressors.

Despite these caveats it turns out that the practice inspired by
the rigorous ideal theory performs quite well. 
We feel this is an example that an {\em ad hoc\/}
approximation guided by a good theory is preferable above 
{\em ad hoc\/} approaches without underlying theoretical foundation.

\subsection{The quartet method}

\begin{figure}[htb]
\begin{center}
\epsfig{file=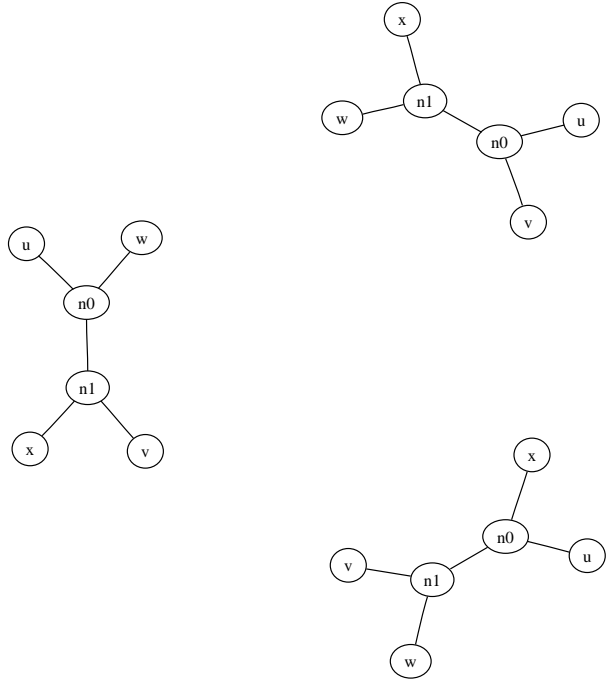,width=8cm}
\end{center}
\caption{The three possible quartets for the set of leaf labels {\em u,v,w,x} }\label{figquart}
\end{figure}

The above approach allows us to compute the distance between 
any pair of objects (any two pieces of music).
We now need to cluster the objects, so that objects that are similar
according to our metric are placed close together. 
We do this by computing a phylogeny tree
based on these distances. Such a phylogeny tree can represent
evolution of species but more widely simply accounts for
closeness of objects from a set with a distance
metric. Such a tree will group objects in subtrees:
the clusters. To find the phylogeny tree there are many methods.
One of the most popular is the quartet method. The idea is as 
follows: we consider every group of four elements from our set
of $n$ elements (in this case, musical pieces); 
there are ${n \choose 4}$ such groups.
From each group $u,v,w,x$ we construct a tree of arity 3,
which implies that the tree consists of two subtrees of two
leaves each. Let us call such a tree a {\em quartet}.  There are
three possibilities denoted (i) $uv | wx$, (ii) $uw | vx$,
and (iii)  $ux | vw$, where a vertical bar divides the two pairs of leaf nodes
into two disjoint subtrees (Figure~\ref{figquart}).

The cost of a quartet is defined as the sum 
of the distances between each pair of neighbors; that
is, $C_{uv|wx} = d(u,v) + d(w,x)$.  For any given tree $T$ and any group
of four leaf labels $u,v,w,x$, we say $T$ is $consistent$ with $uv | wx$
if and only if the path from $u$ to $v$ does not cross
the path from $w$ to $x$.  Note that exactly one of the three possible
quartets for any set of 4 labels must be consistent for any given tree.
\begin{figure}[htb]
\begin{center}
\epsfig{file=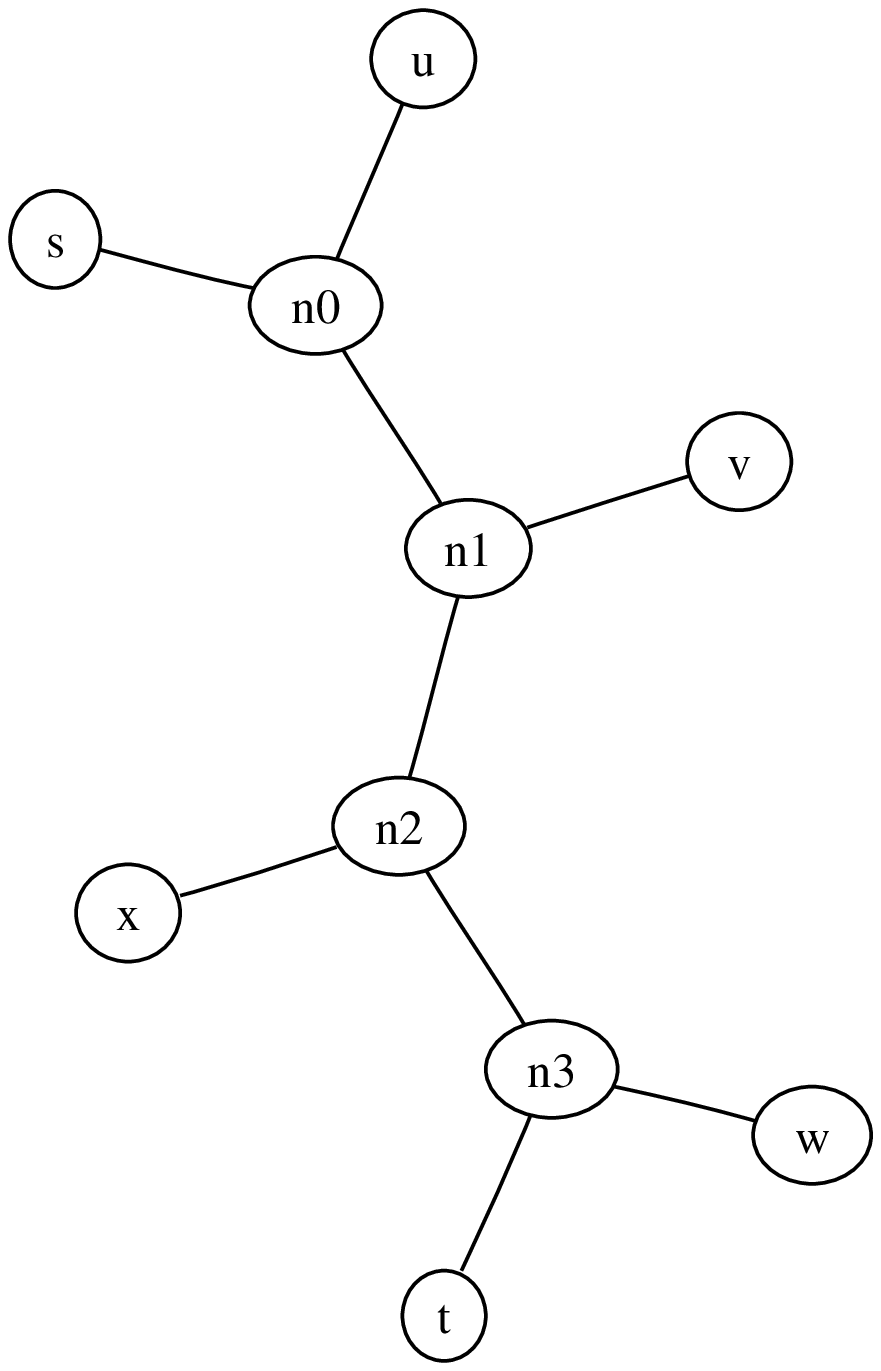,width=5cm}
\end{center}
\caption{An example tree consistent with quartet $uv | wx$ }\label{figquartex}
\end{figure}
We may think of a large tree having many smaller quartet trees embedded
within its structure  (Figure~\ref{figquartex}).  The total cost of a large tree is defined to be the
sum of the costs of all consistent quartets. 
First, generate a list of all possible quartets for all groups of labels
under consideration.  For each group of three possible quartets for a given
set of four labels, calculate a best (minimal) cost, and a worst (maximal)
cost.  Summing all best quartets yields the best (minimal) cost.
Conversely, summing all worst quartets yields the worst (maximal) cost.
The minimal and maximal values need not be attained by actual trees,
however the score of any tree will lie between these two values.
In order to be able to compare tree scores in a more uniform way,
we now rescale the score linearly such that the worst score maps to 0,
and the best score maps to 1, and term this the 
{\em normalized tree benefit score} $S(T)$.
The goal of the quartet method is to find a full tree with a maximum value
of $S(T)$, which is to say, the lowest total cost.
This optimization problem is known to be NP-hard \cite{Ji01} (which means that
it is infeasible in practice) but we can sometimes solve it, and
always approximate it. The current
methods in \cite{Br00} are far too computationally intensive;
they run many months or years on moderate-sized problems
of 30 objects. We have designed a simple method based
on randomization and hill-climbing.  First, a random tree with $2n-2$ nodes
is created, consisting of $n$ leaf nodes (with 1 connecting edge) labeled 
with the names of musical pieces, and $n-2$ non-leaf or {\em internal} nodes
labeled with the lowercase letter ``n'' followed by a unique integer identifier.  Each internal node has exactly three connecting edges.  For this 
tree $T$, we calculate the total cost of all consistent quartets, 
and invert and scale this value to 
find $S(T)$.  Typically, a random tree will be consistent with around
$\frac{1}{3}$ of all quartets.
Now, this tree is denoted the currently best known tree, and is used as
the basis for further searching.  We define a simple mutation on a tree
as one of the three possible transformations:
\begin{enumerate}
\item A {\em leaf swap}, which consists of randomly choosing two leaf nodes
and swapping them.
\item A {\em subtree swap}, which consists of randomly choosing two internal 
nodes and swapping the subtrees rooted at those nodes.
\item A {\em subtree transfer}, whereby a randomly chosen subtree (possibly a leaf) is detached and reattached in another place, maintaining arity invariants.
\end{enumerate}
Each of these simple mutations keeps invariant the
number of leaf and internal nodes in the tree; only the structure and placements
change.  Define a full mutation as a sequence of at least one but potentially
many simple mutations, picked according to the following distribution. 
First we pick the number $k$ of simple mutations that we will perform with
probability $2^{-k}$.  For each such simple mutation, we choose one of
the three types listed above with equal probability.  Finally, for each of 
these simple mutations, we pick leaves or internal nodes, as necessary.  Notice
that trees which are close to the original tree (in terms of number of 
simple mutation steps in between) are examined often, while trees that are 
far away from the original tree will eventually be examined, but not very 
frequently.
So in order to search for a better tree,
we simply apply a full mutation on $T$ to arrive at $T'$, and then
calculate $S(T')$.  If $S(T') > S(T)$, then keep $T'$ as the new best tree.
Otherwise, try a new different tree and repeat.  If $S(T')$ ever reaches
$1$, then halt, outputting the best tree.  Otherwise, run until it seems
no better trees are being found in a reasonable amount of time, in which
case the approximation is complete.

\begin{figure}[htb]
\begin{center}
\epsfig{file=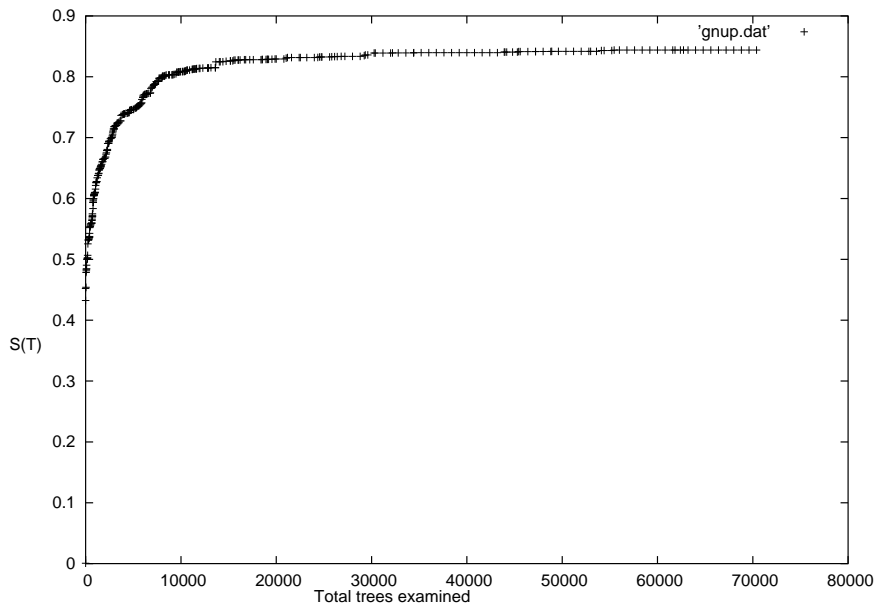,width=8cm,angle=270}
\end{center}
\caption{Progress of the 60-piece experiment over time}\label{figprogress}
\end{figure}

Note that if a tree is ever found such that $S(T) = 1$, then we can stop
because we can be certain that this tree is optimal, as no tree could 
have a lower cost.  In fact, this perfect tree result is achieved in our 
artificial tree reconstruction experiment (Section~\ref{sect.artificial}) 
reliably in less than ten minutes.  For real-world data, $S(T)$ reaches 
a maximum somewhat 
less than $1$, presumably reflecting inconsistency in the distance matrix 
data fed as input to the algorithm, or indicating a search space too large
to solve exactly.
On many typical problems of up to 40 objects this tree-search gives a tree 
with $S(T) \geq 0.9$ within half an hour.  For large numbers of objects,
tree scoring itself can be slow (as this takes order $n^4$ computation steps), 
and the space of
trees is also large, so the algorithm may slow down substantially.
For larger experiments, we use a C++/Ruby implementation with MPI (Message
Passing Interface, a common standard used on massively parallel computers) on a
cluster of workstations in parallel to find trees more rapidly. We can
consider the graph of Figure~\ref{figprogress},
mapping the achieved $S(T)$ score as a function
of the number of trees examined.  Progress
occurs typically in a sigmoidal fashion towards a maximal value $\leq 1$.  

A problem with the outcomes is as follows: For natural
data sets we often see 
some leaf nodes (data items) placed near the center of the tree as singleton
leaves attached to internal nodes, without sibling leaf
nodes.  This results in a more linear, stretched out, and less
balanced, tree. Such trees, even if they represent the underlying
distance matrix faithfully, are hard to fully understand
and may cause misunderstanding of represented relations and clusters.
To counteract this effect, and to bring out the clusters of
related items more visibly, we have added a penalty term of
the following form: For each internal node with exactly one leaf
node attached, the tree's score is reduced by 0.005.  This induces a
tendency in the
algorithm to avoid producing degenerate mostly-linear trees in the
face of data that is somewhat inconsistent, and creates balanced and
more illuminating clusters. It should be noted that the penalty term
causes the algorithm in some cases to settle for a slightly lower
$S(T)$ score than it would have without penalty term. Also the
value of the penalty term is heuristically chosen. The largest 
experiment used 60 items, and we typically had only
a couple of orphans causing a penalty of only a few percent.
This should be set off against the final $S(T)$ score of above 0.85.

Another practicality concerns the stopping criterion, at which $S(T)$
value we stop. Essentially we stopped when the $S(T)$
value didn't change after examining a large number of mutated trees. 
An example is the progress of Figure~\ref{figprogress},

\section{Details of Our Implementation}\label{secdetails}

Initially, we downloaded 118 separate MIDI (Musical Instrument Digital
Interface, a versatile digital music format
available on the world-wide-web) 
files selected from a range of classical composers, as well as some
popular music.   Each of these files was run through a preprocessor 
to extract just MIDI Note-On
and Note-Off events.  These events were then converted to a player-piano
style representation, with time quantized in $0.05$ second intervals.
All instrument indicators, MIDI Control signals, and tempo variations were 
ignored.  For each track in the MIDI file, we calculate two quantities:
An {\em average volume} and a {\em modal note}.
The average volume is calculated by averaging the volume (MIDI Note velocity)
of all notes in the track.  The modal note is defined to be the note 
pitch that sounds most often in that track.  If this is not unique, 
then the lowest such note is chosen.  The modal note is used as a 
key-invariant reference point from which to represent all notes.  
It is denoted by $0$, higher notes are denoted by positive numbers, and 
lower notes are denoted by negative numbers.  A value of $1$ indicates 
a half-step above the modal note, and a value of $-2$ indicates
a whole-step below the modal note.  The tracks are sorted according to
decreasing average volume, and then output in succession.  For each track,
we iterate through each time sample in order, outputting a single signed
8-bit value for each currently sounding note.  Two special values are
reserved to represent the end of a time step and the end of a track.  This
file is then used as input to the compression stage for distance
matrix calculation and subsequent tree search.

\section{Results}\label{secresults}

\subsection{Three controlled experiments}\label{sect.artificial}

With the natural data sets of music pieces that we use, one may have the preconception 
(or prejudice) that music by Bach should be clustered together, 
music by Chopin should be clustered together, and so should music by
rock stars. However, the preprocessed music files of a piece by Bach and
a piece by Chopin, or the Beatles, may resemble one another 
more than two different
pieces by Bach---by accident or indeed by design and copying. Thus, natural
data sets may have ambiguous, conflicting, or counterintuitive 
outcomes. In other words, the experiments on actual pieces have
the drawback of not having one clear ``correct'' answer that can 
function as a benchmark for assessing our experimental outcomes.
Before describing the experiments we did with MIDI files of actual 
music, we discuss three experiments that show that our
program indeed does what it is supposed to do---at least in 
artificial situations where we know in advance what the correct answer is.
The similarity machine consists of two parts: (i) extracting a distance matrix
from the data, and (ii) constructing a tree 
from the distance matrix using our novel quartet-based heuristic.

\begin{figure}[htb]
\begin{center}
\epsfig{file=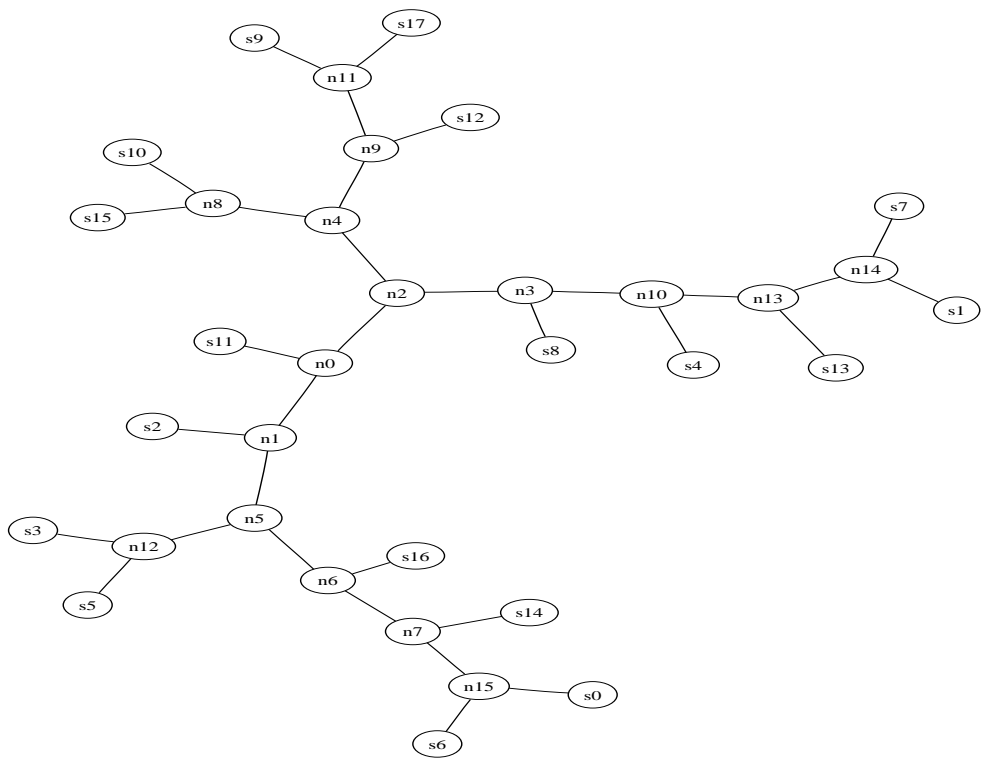,width=13cm,height=10cm}
\end{center}
\caption{The tree that our algorithm reconstructed}\label{figarttreereal}
\end{figure}

{\bf Testing the quartet-based tree construction:}
We first test whether the quartet-based tree construction
heuristic is trustworthy: 
We generated a random ternary tree $T$ with 18 leaves, and derived
a distance metric from it by defining the distance between 
two nodes as follows:
Given the length of the path from $a$ to $b$, in an integer number of
edges, as $L(a,b)$, let 
\[d(a,b) = { {L(a,b)+1} \over 18},
\]
  except when
$a = b$, in which case $d(a,b) = 0$.  It is easy to verify that this
simple formula always gives a number between 0 and 1, and is monotonic
with path length.
Given only the $18\times 18$ matrix of these normalized distances, 
our quartet method exactly reconstructed $T$ represented in
Figure~\ref{figarttreereal}, with $S(T)=1$.

\begin{figure}[htb]
\begin{center}
\epsfig{file=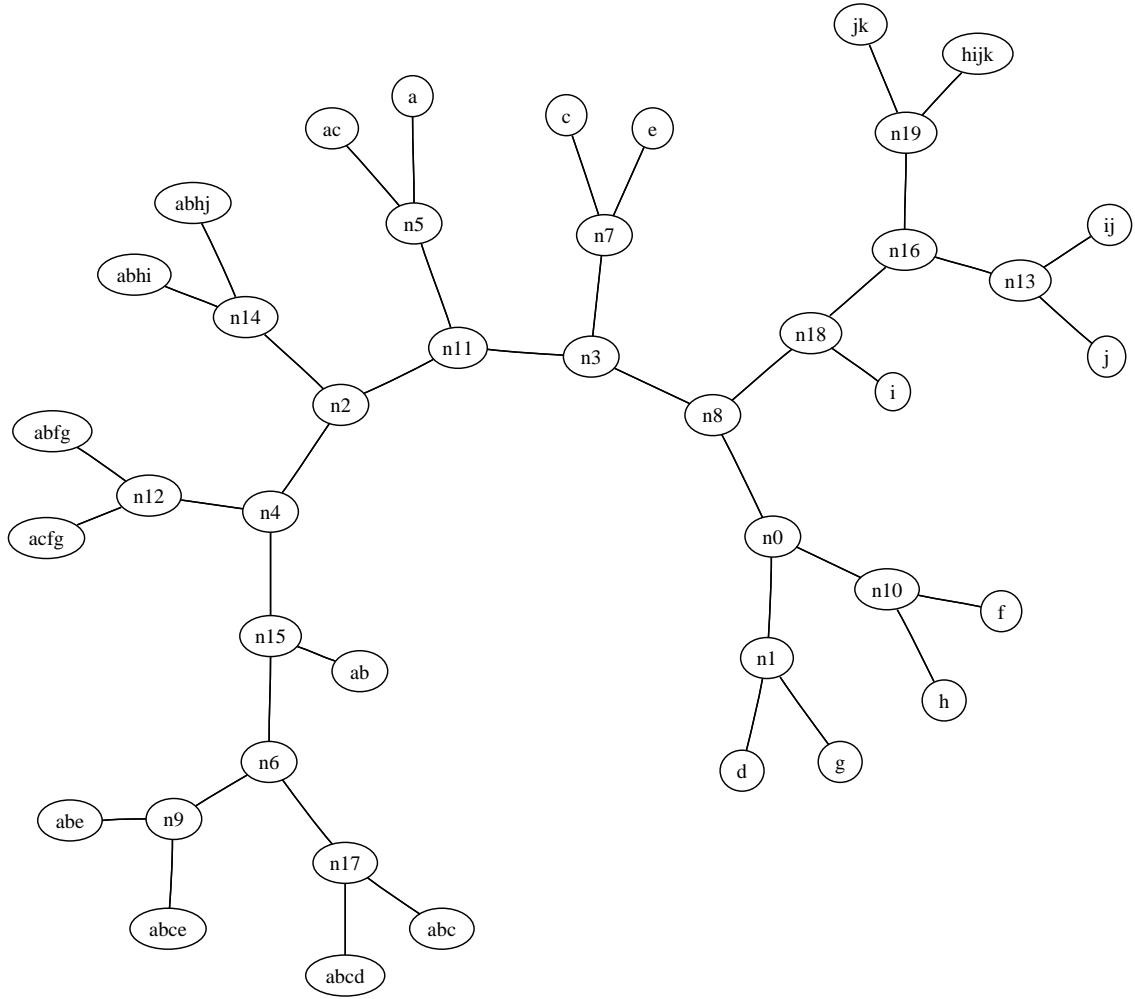,width=15cm}
\end{center}
\caption{Classification of artificial files with repeated 1-kilobyte tags }\label{figtaggedfiles}
\end{figure}

{\bf Testing the similarity machine on artificial data:}
Given that the tree reconstruction method is accurate
on clean consistent data, we tried whether the full procedure
works in an acceptable manner when we know what the outcome should
be like:
\begin{figure}[htb]
\begin{center}
\epsfig{file=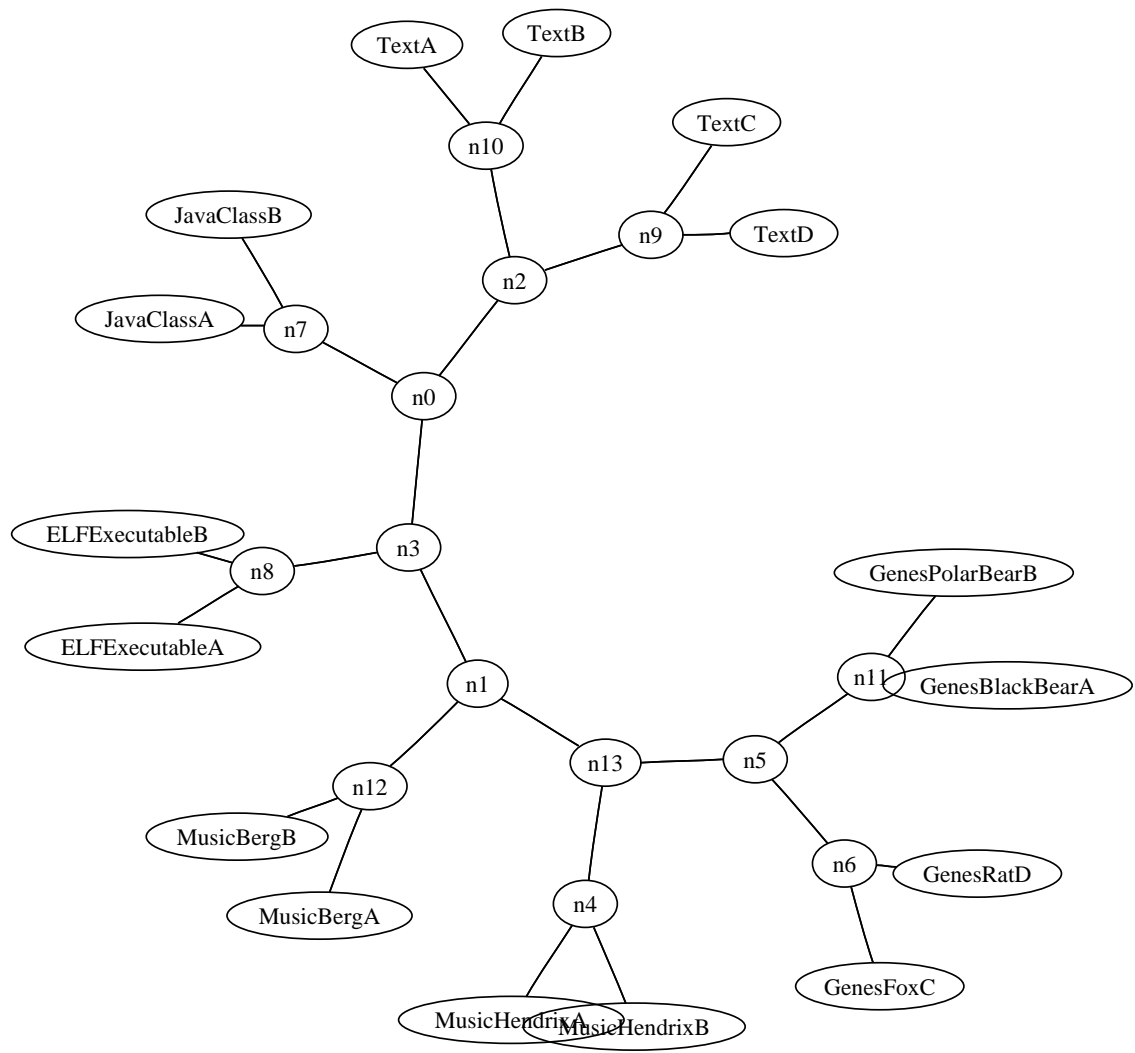,width=15cm}
\end{center}
\caption{Classification of different file types}\label{figfiletypes}
\end{figure}
We randomly generated 22 separate 1-kilobyte blocks of data where
each byte was equally probable and called these {\em tags}.  Each tag
was associated with a different lowercase letter of the alphabet.  Next,
we generated 80-kilobyte files by starting with a block of purely random
bytes and applying one, two, three, or four different tags on it.
Applying a tag consists of ten repetitions of picking a random location
in the 80-kilobyte file, and overwriting that location with the universally
consistent tag that is indicated.  So, for instance, to create the file
referred to in the diagram by ``a'', we start with 80 kilobytes of random data,
then pick ten places to copy over this random data with the arbitrary 
1-kilobyte sequence identified as tag {\em a}.  Similarly, to create file ``ab'',
we start with 80 kilobytes of random data, then pick ten places to put
copies of tag {\em a}, then pick ten more places to put copies of tag {\em b} (perhaps
overwriting some of the {\em a} tags).  Because we never use more than four
different tags, and therefore never place more than 40 copies of tags, we
can expect that at least half of the data in each file is random and
uncorrelated with the rest of the files.  The rest of the file is 
correlated with other files that also contain tags in common; the more 
tags in common, the more related the files are.
The resulting tree is given in Figure~\ref{figtaggedfiles}; it can be
seen that clustering occurs exactly as we would expect.
The $S(T)$ score is 0.905.

{\bf Testing the similarity machine on natural data:}
We test gross classification of files
based on markedly different file types.  Here, we chose several files:
\begin{enumerate}
\item Four mitochondrial gene sequences, from a black bear, polar bear, 
fox, and rat.
\item Four excerpts from the novel { \em The Zeppelin's Passenger} by 
E.~Phillips Oppenheim
\item Four MIDI files without further processing; two from Jimi Hendrix and 
two movements from Debussy's Suite bergamasque
\item Two Linux x86 ELF executables (the {\em cp} and {\em rm} commands)
\item Two compiled Java class files.
\end{enumerate}
As expected, the program correctly classifies each of the different types
of files together with like near like. The result is reported
in Figure~\ref{figfiletypes} with $S(T)$ equal to 0.984.

\subsection{Genres: rock vs.~jazz vs.~classical}

\begin{figure}[htb]
\begin{center}
\epsfig{file=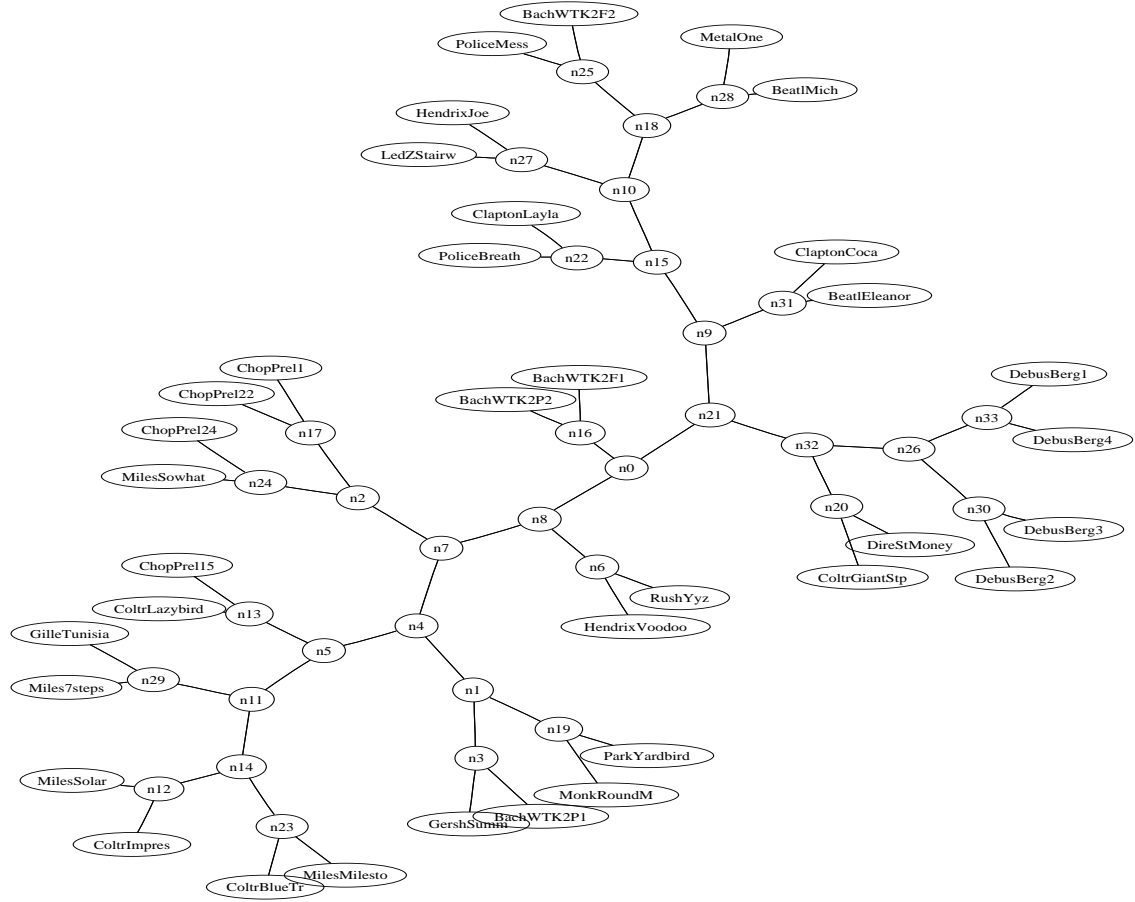,width=15cm,height=12cm}
\end{center}
\caption{Output for the 36 pieces from 3 genres}\label{figgenres}
\end{figure}

Before testing whether our program can see the distinctions
between various classical composers, we first 
show that it can distinguish between three broader musical genres:
classical music, rock, and jazz. This should be easier than
making distinctions ``within'' classical music. 
All musical pieces we used are listed in the tables in the appendix.
For the genre-experiment we used 12 classical pieces (the small set 
from Table~\ref{tableclassicalpieces}, consisting of Bach, Chopin, and Debussy),
12 jazz pieces (Table~\ref{tablejazzpieces}), and
12 rock pieces (Table~\ref{tablerockpieces}).
The tree that our program came up with is given in Figure~\ref{figgenres}.
The $S(T)$ score is 0.858.

The discrimination between the 3 genres is good but not perfect.
The upper branch of the tree contains 10 of the 12 jazz pieces,
but also Chopin's Pr\'elude no.~15 and a Bach Prelude.
The two other jazz pieces, Miles Davis' ``So what'' and John 
Coltrane's ``Giant steps'' are placed elsewhere in the tree,
perhaps according to some kinship that now escapes us but can be
identified by closer studying of the objects concerned.
Of the rock pieces, 9 are placed close together in the rightmost branch, 
while Hendrix's ``Voodoo chile'', Rush' ``Yyz'', 
and Dire Straits' ``Money for nothing'' are further away.
In the case of the Hendrix piece this may be explained by the fact
that it does not fit well in a specific genre.
Most of the classical pieces are in the lower left part of the tree.
Surprisingly, 2 of the 4 Bach pieces are placed elsewhere.
It is not clear why this happens and may be considered an error
of our program, since we perceive the 4 Bach pieces to  be very close,
both structurally and melodically (as they all come from the mono-thematic
``Wohltemperierte Klavier'').
However, Bach's is a seminal music and has been copied and cannibalized
in all kinds of recognizable or hidden manners; closer scrutiny could
reveal likenesses in its present company that are not now apparent to us.
In effect our similarity engine aims at the ideal of a perfect
data mining process, discovering unknown features in which the
data can be similar. 

\subsection{Classical piano music (small set)}

\begin{figure}
\begin{center}
\epsfig{file=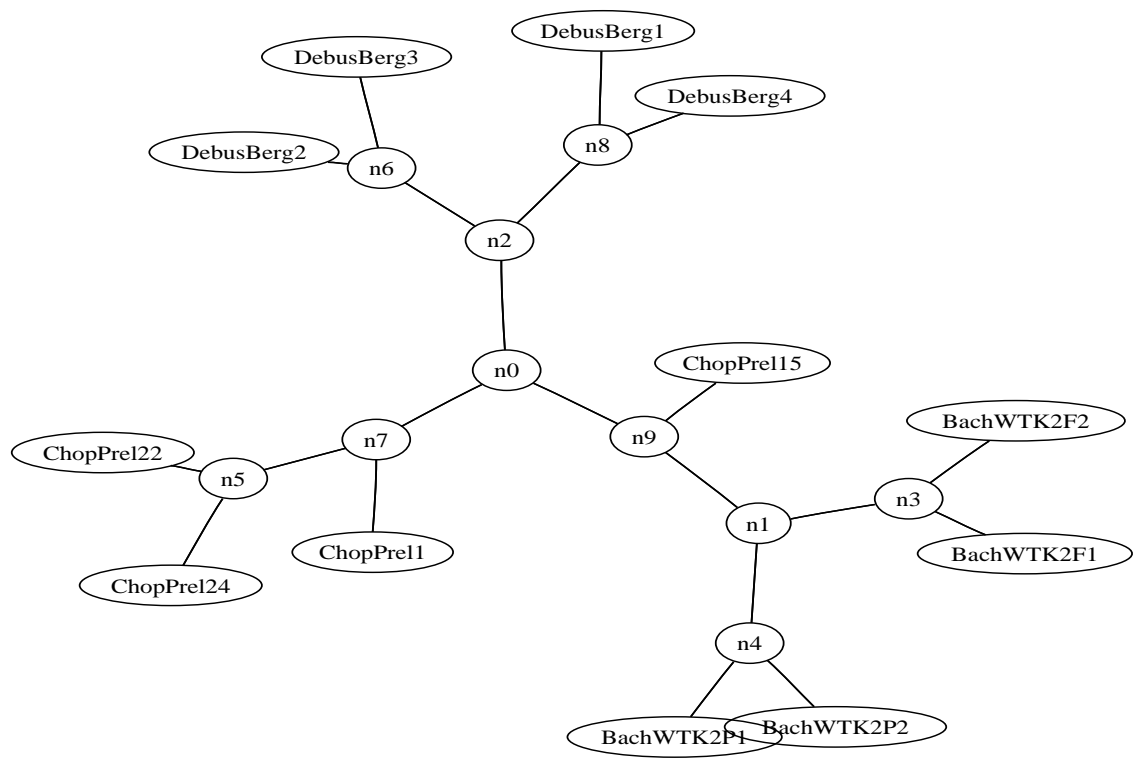,width=15cm,height=10cm}
\end{center}
\caption{Output for the 12-piece set}\label{figsmallset}
\end{figure}

In Table~\ref{tableclassicalpieces} we list all 60 classical piano pieces used, 
together with their abbreviations. Some of these are complete
compositions, others are individual movements from larger compositions.
They all are piano pieces, but experiments on 34 movements of symphonies
gave very similar results (Section~\ref{secsymphonies}).
Apart from running our program on the whole set of 60 piano 
pieces, we also tried it on two smaller sets: a small 12-piece set, 
indicated by `(s)' in the table, and a medium-size 32-piece set, 
indicated by `(s)' or `(m)'.

The small set encompasses the 4 movements from Debussy's Suite bergamasque,
4 movements of book 2 of Bach's Wohltemperierte Klavier, and 4 preludes from 
Chopin's opus~28. As one can see in Figure~\ref{figsmallset}, 
our program does a pretty good job at clustering these pieces.
The $S(T)$ score is also high: 0.958.
The 4 Debussy movements form one cluster, as do the 4 Bach pieces.  
The only imperfection in the tree, judged by what one would 
intuitively expect, is that Chopin's Pr\'elude no.~15 lies a bit closer 
to Bach than to the other 3 Chopin pieces.
This Pr\'elude no~15, in fact, consistently forms an odd-one-out 
in our other experiments as well. This is an example of pure data mining,
since there is some musical truth 
to this, as no.~15 is perceived as by far the most eccentric 
among the 24 Pr\'eludes of Chopin's opus~28.

\subsection{Classical piano music (medium set)}
\begin{figure}[hbt]
\begin{center}
\epsfig{file=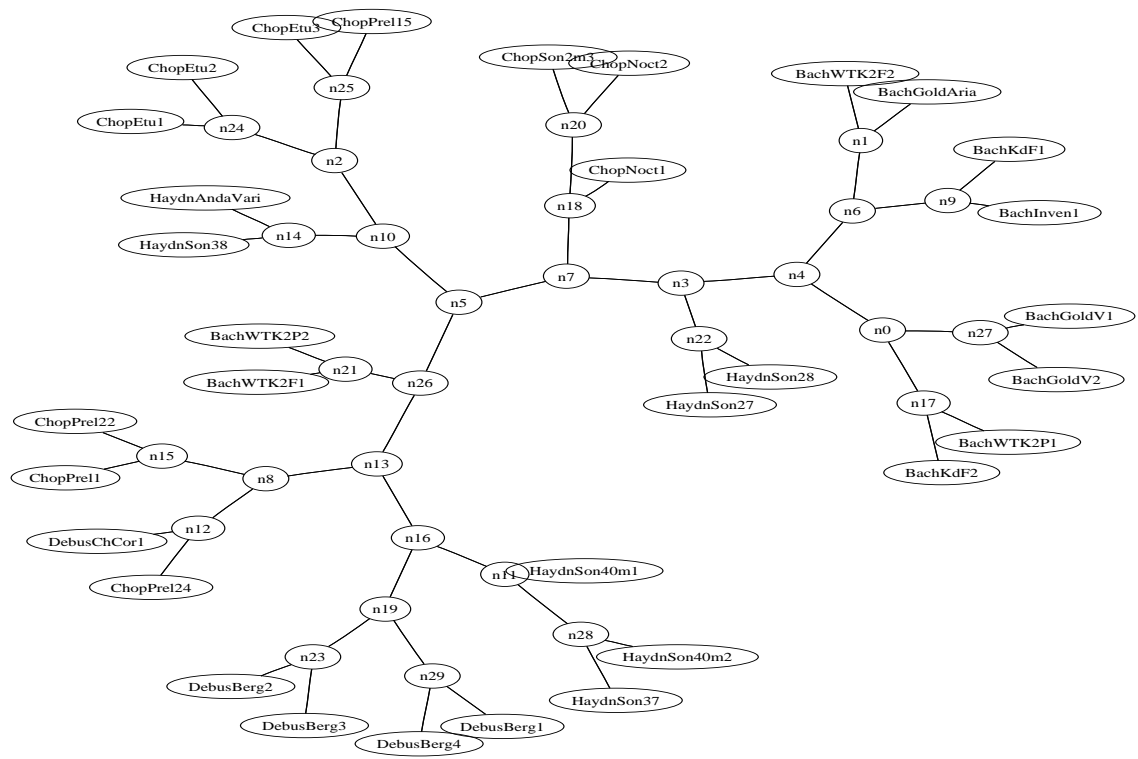,width=15cm,height=10cm}
\end{center}
\caption{Output for the 32-piece set}\label{figmediumset}   
\end{figure}

The medium set adds 20 pieces to the small set:
6 additional Bach pieces, 6 additional Chopins, 1 more Debussy piece,
and 7 pieces by Haydn. The experimental results are given in
Figure~\ref{figmediumset}. The $S(T)$ score is slightly lower than
in the small set experiment: 0.895.
Again, there is a lot of structure
and expected clustering. Most of the Bach pieces are together,
as are the four Debussy pieces from the Suite bergamasque.
These four should be together because they are movements from the same piece;
The fifth Debussy item is somewhat apart since it comes from another piece.
Both the Haydn and the Chopin
pieces are clustered in little sub-clusters of two or three pieces, 
but those sub-clusters are scattered throughout the tree instead
of being close together in a larger cluster.
These small clusters may be an imperfection of the method,
or, alternatively point at musical similarities between the
clustered pieces that transcend the similarities induced by
the same composer. Indeed, this may point the way for further
musicological investigation.

\subsection{Classical piano music (large set)}

\begin{figure}
\begin{center}
\epsfig{file=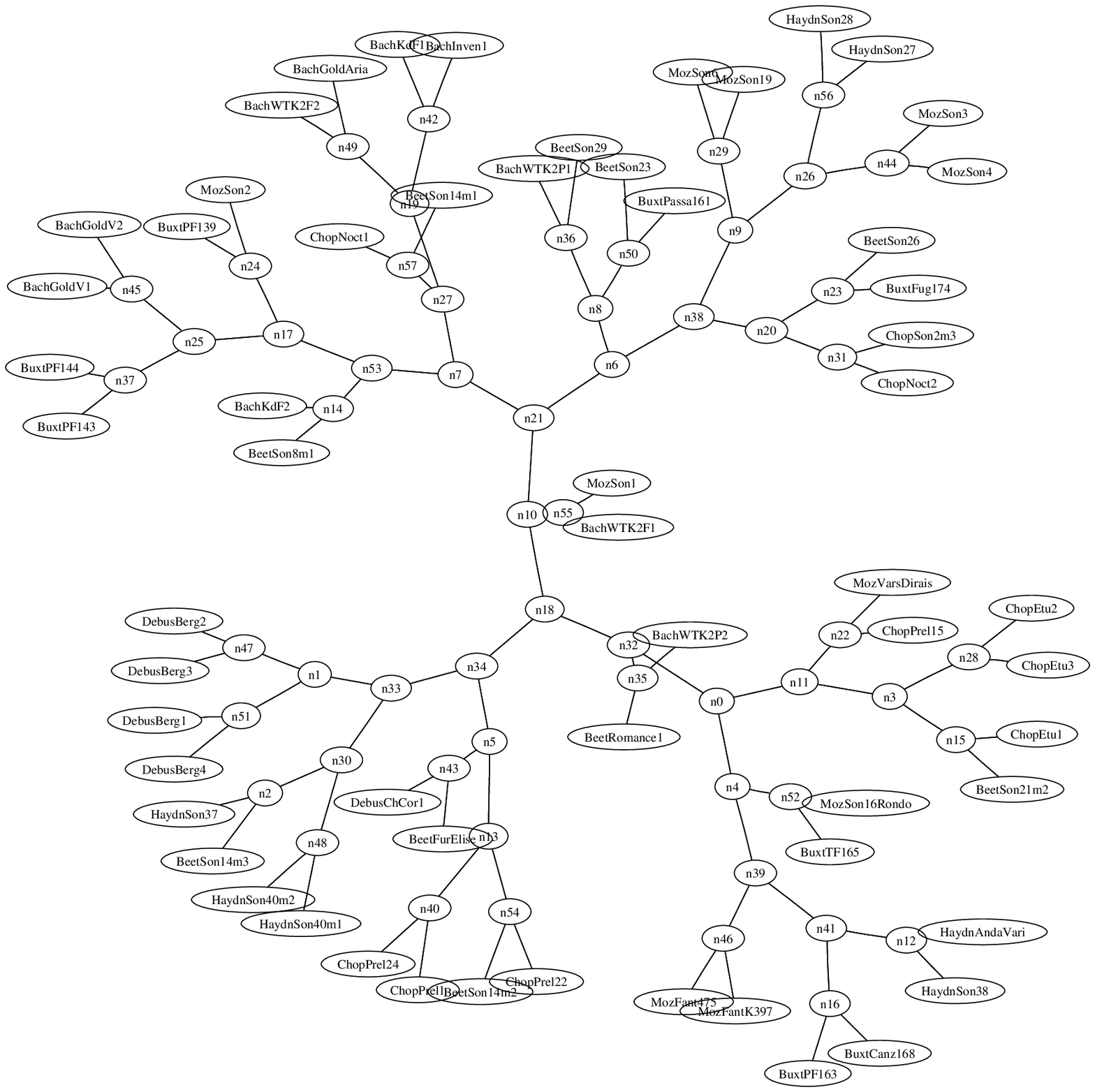,width=17cm,height=10cm}
\end{center}
\caption{Output for the 60-piece set}\label{figlargeset}   
\end{figure}

Figure~\ref{figlargeset} gives the output of a run of our program 
on the full set of 60 pieces. This adds 10 pieces by Beethoven, 
8 by Buxtehude, and 10 by Mozart to the medium set.
The experimental results are given in Figure~\ref{figlargeset}. 
The results are still far from random, but leave more to 
be desired than the smaller-scale experiments.
Indeed, the $S(T)$ score has dropped further from that of the
medium-sized set to 0.844.
This may be an artifact of the interplay between the relatively small
size, and large number, of the files compared: (i) the distances
estimated are less accurate; (ii) the number of quartets
with conflicting requirements increases; and (iii) the computation
time rises to such an extent that the correctness score of the
displayed cluster graph within the set time limit
is lower than in the smaller samples. 
Nonetheless, Bach and Debussy are still reasonably well clustered,
but other pieces (notably the Beethoven and Chopin ones)
are scattered throughout the tree. Maybe this means
that individual music pieces by these composers are more similar
to pieces of other composers than they are to each other?
The placement of the pieces is closer to intuition on a small level
(for example, most pairing of siblings corresponds to musical similarity in
the sense of the same composer) than on the larger level.
This is similar to the phenomenon of little sub-clusters 
of Haydn or Chopin pieces that we saw in the medium-size experiment.

\subsection{Clustering symphonies}\label{secsymphonies}

\begin{figure}
\begin{center}
\epsfig{file=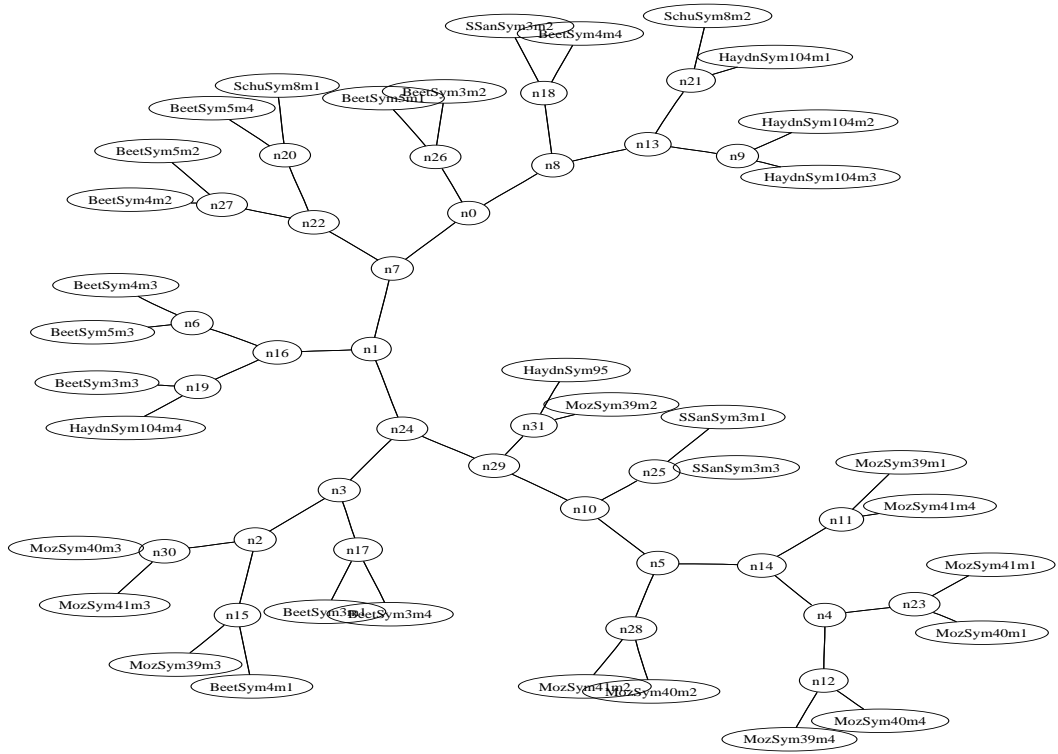,width=14cm,height=10cm}
\end{center}
\caption{Output for the set of 34 movements of symphonies}\label{figsymphonies}
\end{figure}

Finally, we tested whether the method worked for more complicated
music, namely 34 symphonic pieces.  We took
two Haydn symphonies (no.~95 in one file, and the four movements of~104), 
three Mozart symphonies (39, 40, 41), 
three Beethoven symphonies (3, 4, 5),
of Schubert's Unfinished symphony, and of Saint-Saens Symphony no.~3.
The results are reported in Figure~\ref{figsymphonies},
with a quite reasonable $S(T)$ score of 0.860.

\section{Future Work and Conclusion}

Our research raises many questions worth looking into further:
\begin{itemize}
\item The program can be used as a data mining machine to discover
hitherto unknown similarities between music pieces of different
composers or indeed different genres. In this manner we can discover
plagiarism or indeed honest influences between music pieces and
composers. Indeed, it is thinkable that we can use the method
to discover seminality of composers, or separate music eras
and fads.
\item A very interesting application of our program
would be to select a plausible composer for a newly 
discovered piece of music of which the composer is not known.
In addition to such a piece, this experiment would 
require a number of pieces from known composers that 
are plausible candidates.  We would just run our program 
on the set of all those pieces, and see where the new 
piece is placed.  If it lies squarely within a cluster 
of pieces by composer such-and-such, then that would be 
a plausible candidate composer for the new piece.
\item Each run of our program is different---even on 
the same set of data---because of our use of randomness for 
choosing mutations in the quartet method. 
It would be interesting to investigate more precisely 
how stable the outcomes are over different such runs.
\item At various points in our program, somewhat 
arbitrary choices were made.
Examples are the compression algorithms we use
(all practical compression algorithms will fall short
of Kolmogorov complexity, but some less so than others);
the way we transform the MIDI files (choice of length 
of time interval, choice of note-representation);
the cost function in the quartet method.
Other choices are possible and may or may not lead 
to better clustering.\footnote{We compared the quartet-based
approach to the tree reconstruction 
with alternatives.  One such alternative that we tried is to
compute the Minimum Spanning Tree (MST) from the matrix
of distances. MST has the advantage of being very efficiently 
computable, but resulted in trees that were much worse than
the quartet method. It appears that the quartet method is extremely sensitive
in extracting information even from small differences in the entries
of the distance matrix, where other methods would be led to error.}
Ideally, one would like to have well-founded theoretical 
reasons to decide such choices in an optimal way. 
Lacking those, trial-and-error seems the only way 
to deal with them.
\item The experimental results got decidedly worse when
the number of pieces grew.
Better compression methods may improve this situation, but the effect
is probably due to unknown scaling problems with the quartet
method or nonlinear scaling of possible similarities in a larger
group of objects (akin to the phenomenon described in the so-called
``birthday paradox'': in a group of about two dozen people there
is a high chance that at least two of the people have the
same birthday). Inspection of the underlying distance matrices
makes us suspect the latter.
\item Our program is not very good at dealing with very 
small data files (100 bytes or so), because significant
compression only kicks in for larger files.
We might deal with this by comparing various sets of such 
pieces against each other, instead of individual ones.
\end{itemize}

\subsection*{Acknowledgments}
We thank John Tromp for useful discussions.

\appendix

\section{Appendix: The Music Pieces Used}
\begin{table}[htb]
\begin{center}
\begin{tabular}{|l|l|l|} \hline
Composer & Piece & Acronym\\ \hline\hline
J.S.~Bach (10) & Wohltemperierte Klavier II: Preludes and fugues 1,2 & BachWTK2\{F,P\}\{1,2\} (s)\\
          & Goldberg Variations: Aria, Variations 1,2  & BachGold\{Aria,V1,V2\} (m) \\
          & Kunst der Fuge: Variations 1,2             & BachKdF\{1,2\} (m) \\
          & Invention 1                                & BachInven1 (m) \\
Beethoven (10) & Sonata no.~8 (Pathetique), 1st movement    & BeetSon8m1 \\ 
          & Sonata no.~14 (Mondschein), 3 movements    & BeetSon14m\{1,2,3\}\\
          & Sonata no.~21 (Waldstein), 2nd movement    & BeetSon21m2\\
          & Sonata no.~23 (Appassionata)               & BeetSon23\\
          & Sonata no.~26 (Les Adieux)                 & BeetSon26\\
          & Sonata no.~29 (Hammerklavier)              & BeetSon29\\
          & Romance no.~1                              & BeetRomance1\\
          & F\"ur Elise                                & BeetFurElise\\
Buxtehude (8) & Prelude and fugues, BuxWV 139,143,144,163  & BuxtPF\{139,143,144,163\} \\
          & Toccata and fugue, BuxWV 165               & BuxtTF165 \\
          & Fugue, BuxWV 174                           & BuxtFug174\\
          & Passacaglia, BuxWV 161                     & BuxtPassa161\\
          & Canzonetta, BuxWV 168                      & BuxtCanz168\\
Chopin (10) & Pr\'eludes op.~28: 1, 15, 22, 24 & ChopPrel\{1,15,22,24\} (s)\\
          & Etudes op.~10, nos.~1, 2, and 3            & ChopEtu\{1,2,3\} (m)\\
          & Nocturnes nos.~1 and 2                     & ChopNoct\{1,2\} (m)\\
          & Sonata no.~2, 3rd movement                 & ChopSon2m3 (m)\\
Debussy (5) & Suite bergamasque, 4 movements             & DebusBerg\{1,2,3,4\} (s)\\
          & Children's corner suite (Gradus ad Parnassum) & DebusChCorm1 (m)\\
Haydn (7) & Sonatas nos.~27, 28, 37, and 38            & HaydnSon\{27,28,37,38\} (m)\\
          & Sonata no.~40, movements 1,2               & HaydnSon40m\{1,2\} (m)\\
          & Andante and variations                     & HaydnAndaVari (m)\\
Mozart (10) & Sonatas nos.~1,2,3,4,6,19                & MozSon\{1,2,3,4,6,19\} \\
          & Rondo from Sonata no.~16                   & MozSon16Rondo \\
          & Fantasias K397, 475                        & MozFantK\{397,475\} \\
          & Variations ``Ah, vous dirais-je madam''    & MozVarsDirais\\ \hline
\end{tabular}
\end{center}
\caption{The 60 classical pieces used 
(`m' indicates presence in the medium set, `s' in the small and medium sets)}\label{tableclassicalpieces}
\end{table}

\begin{table}[htb]
\begin{center}
\begin{tabular}{|l|l|} \hline
John Coltrane& Blue trane\\
             & Giant steps\\
             & Lazy bird\\
             & Impressions\\
Miles Davis  & Milestones\\
             & Seven steps to heaven\\ 
             & Solar\\
             & So what\\
George Gershwin & Summertime\\
Dizzy Gillespie & Night in Tunisia\\
Thelonious Monk & Round midnight\\
Charlie Parker  & Yardbird suite\\ \hline
\end{tabular}
\end{center}
\caption{The 12 jazz pieces used}\label{tablejazzpieces}
\end{table}

\begin{table}[htb]
\begin{center}
\begin{tabular}{|l|l|} \hline
The Beatles  & Eleanor Rigby\\
             & Michelle\\
Eric Clapton & Cocaine\\
             & Layla\\
Dire Straits & Money for nothing\\
Led Zeppelin & Stairway to heaven\\
Metallica    & One\\
Jimi Hendrix & Hey Joe\\
             & Voodoo chile\\
The Police   & Every breath you take\\
             & Message in a bottle\\
Rush         & Yyz\\ \hline
\end{tabular}
\end{center}
\caption{The 12 rock pieces used}\label{tablerockpieces}
\end{table}

\end{document}